\documentclass[journal]{IEEEtran}
\usepackage{amsmath,amsfonts}
\usepackage{algorithmic}
\usepackage{array}
\usepackage{cite}
\usepackage[caption=false,font=footnotesize,labelfont=rm,textfont=rm]{subfig}
\usepackage{textcomp}
\usepackage{stfloats}
\usepackage{url}
\usepackage{CJKutf8}
\usepackage{verbatim}
\usepackage{graphicx}
\usepackage{hyperref}
\def\BibTeX{{\rm B\kern-.05em{\sc i\kern-.025em b}\kern-.08em
    T\kern-.1667em\lower.7ex\hbox{E}\kern-.125emX}}
\usepackage{balance}

\begin{document}
\title{Waveform Design for Joint Communication and SAR Imaging Under Random Signaling}
\author{Bowen Zheng and Fan Liu, ~\IEEEmembership{Senior Member,~IEEE}  

\thanks{This work was supported in part by Special Funds for the Cultivation of Guangdong College Students' Scientific and Technological Innovation (``Climbing Program" Special Funds, No. pdjh2024c11603), in part by National Natural Science Foundation of China (NSFC) Project No. 62101234. (Corresponding author: Fan Liu.)

Bowen Zheng and Fan Liu are with the School of System Design and Intelligent Manufacturing (SDIM), Southern University of Science and Technology, Shenzhen 518055, China (e-mail: zhengbw2022@mail.sustech.edu.cn;
liuf6@sustech.edu.cn).
}
}


\maketitle

\begin{abstract}
Conventional synthetic aperture radar (SAR) imaging systems typically employ deterministic signal designs, which lack the capability to convey communication information and are thus not suitable for integrated sensing and communication (ISAC) scenarios. In this letter, we propose a joint communication and SAR imaging (JCASAR) system based on orthogonal frequency-division multiplexing (OFDM) signal with cyclic prefix (CP), which is capable of reconstructing the target profile while serving a communication user. In contrast to traditional matched filters, we propose a least squares (LS) estimator for range profiling. Then the SAR image is obtained followed by range cell migration correction (RCMC) and azimuth processing. By minimizing the mean squared error (MSE) of the proposed LS estimator, we investigate the optimal waveform design for SAR imaging, and JCASAR under random signaling, where power allocation strategies are conceived for Gaussian-distributed ISAC signals, in an effort to strike a flexible performance tradeoff between the communication and SAR imaging tasks. Numerical results are provided to validate the effectiveness of the proposed ISAC waveform design for JCASAR systems.
\end{abstract}

\begin{IEEEkeywords}
Integrated sensing and communication (ISAC), synthetic aperture radar (SAR) imaging, orthogonal frequency-division multiplexing (OFDM), Gaussian signaling.
\end{IEEEkeywords}

\section{Introduction}
\IEEEPARstart{R}{ecently}, the ITU-R has endorsed the draft recommendation for IMT-2030 (6G), incorporating ISAC as one of the six primary application scenarios within the 6G framework \cite{b1}. The fundamental vision of ISAC is to involve the pervasive deployment of wireless sensing capabilities across the 6G network \cite{b2},\cite{b2_1}. This vision is anticipated to serve as a crucial facilitator for many use cases \cite{b3}, among which RF imaging is indispensable, thanks to its ability to generate high-resolution images under diverse conditions, offering all-weather and day-and-night imaging capabilities. By equipping the 6G network with the RF imaging functionality, it may play a critical role in numerous emerging scenarios requiring simultaneous communication and imaging services, such as low-altitude space economy applications, where the UAVs perform environmental monitoring while communicating with users.

The RF imaging systems deployed on UAV platforms typically operate in the SAR mode \cite{b4}. Conventional SAR imaging employs linear frequency modulation or step frequency waveforms. However, in high-bandwidth radar systems, these signals encounter Inter-Range Cell Interference (IRCI), which arises from the presence of numerous range cells within a range line, where sidelobes induced by matched filters disrupt adjacent range cells. Inspired by the orthogonal frequency-division multiplexing (OFDM) technique, a pivotal technology in WiFi (IEEE 802.11), LTE and 5G, OFDM SAR systems have received significant attention in recent years. The investigation into range ambiguity in OFDM SAR was explored in \cite{b5}, while \cite{b6} delved into cross-range reconstruction. Notably, these studies employed imaging algorithms akin to conventional SAR systems. Drawing inspiration from wireless communication applications where an OFDM signal with sufficient cyclic prefix can transform an Inter-Symbol Interference (ISI) channel into multiple ISI-free subchannels, a pioneering algorithm for SAR imaging with CP-OFDM signal was introduced in \cite{b7}. Another innovative work, by reformulating the system to a linear model, contributed to improving the performance of the SAR imaging system \cite{b8}. Moreover, investigations into SAR imaging algorithms involving MIMO-OFDM systems and waveform design were conducted in \cite{b9} and \cite{b10}. However, the OFDM SAR imaging system waveforms proposed in \cite{b5}-\cite{b10} are solely designed for SAR imaging and lack the capability to convey useful information, making them unsuitable in ISAC scenarios.

Recently, there have been noteworthy studies focusing on joint communication and SAR imaging. In \cite{b11}, a waveform design approach based on time-frequency spectrum shaping is proposed to achieve JCASAR. Furthermore, \cite{b12} introduces a comprehensive watermarking framework for JCASAR system. These works generally employ orthogonal resource allocation for achieving JCASAR, where the communication and sensing waveforms are separated over frequency or time domains. However, they are generally unable to optimize the resource utilization to its fullest potential compared to fully unified waveform design \cite{b3}. To realize JCASAR through the latter technique, two primary challenges emerge. Firstly, the signal should possess randomness to effectively convey information, which may degrade the SAR imaging performance. Secondly, the signal should be able to achive a favorable performance tradeoff between imaging and communication. These challenges motivate the study of this letter.

To tackle the above challenges, this letter proposes a novel waveform design approach for JCASAR system by taking into account the data randomness. We begin by introducing the system model and corresponding performance metric. Subsequently, by imposing the MSE of the LS estimator as an objective function, while ensuring the communication achievable rate, we propose an optimal power allocation strategy that achieves a scalable performance tradeoff between imaging and communications. Finally, we verify our analysis through numerical simulation results.

\begin{figure} [t]
\centering
\includegraphics[width=0.4\textwidth]{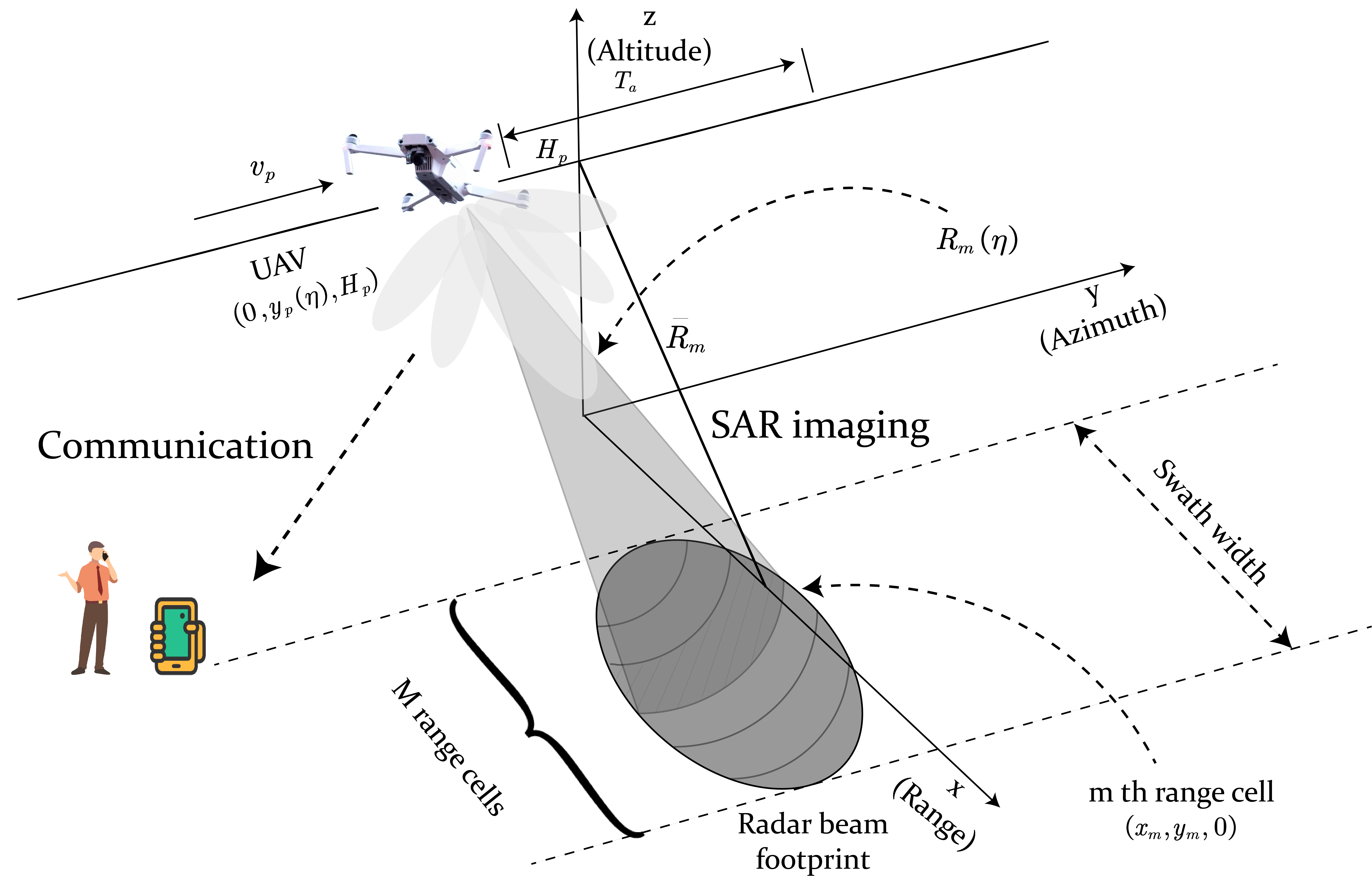}
\caption{Monostatic JCASAR geometry.}
\label{Fig1_geometry}
\end{figure}

\section{System Model}
We consider a UAV for low-altitude space economy applications, which is reconstructing the profile of the target or environment and transmitting the imaging information to the user for navigation or path planning purposes, as illustrated in Fig. \ref{Fig1_geometry}. We assume that the ISAC signal emitted by UAV is unknown to the communication user but is perfectly known at UAV for imaging application. Subsequently, the next two subsections will provide details of the communication model and the SAR imaging model.

\subsection{OFDM Communication Model}
Consider an OFDM signal with $N$ subcarriers, with the subcarrier spacing, bandwidth, and  OFDM symbol duration being given as $\Delta f$, $B=N\varDelta f$, and $T=1/\varDelta f$. Let $S_k$ represents the symbol modulated on the $k$-th subcarrier, and $\sum_{k=0}^{N-1}{\mathbb{E}[|S_k|^2]}=N$.

The discrete-time OFDM signal can be represented as the inverse fast Fourier transform (IFFT) of the vector
\begin{equation}
s(t) =\frac{1}{\sqrt{N}}\sum_{k=0}^{N-1}{S_k\exp \left\{ j2\pi k\varDelta ft \right\}}, t\in[ 0,T+T_{CP}],
\label{transimtted signal}
\end{equation}
the symbol $ T_{CP} $ represent the time duration of the cyclic prefix, which aims to transform an Inter-Symbol Interference (ISI) channel into multiple ISI-free subchannels. 

In this scenario, the channel between UAV and the communication user mainly depends on pass loss effect (i.e. $P_L=-10\log _{10}\frac{G_l\lambda ^2}{\left( 4\pi d \right) ^2}$ (dB), where $G_l$ is the product of the transmit and receive antenna gains), which is up to the distance between UAV and the communication user. Due to the fact that the UAV may be trated as quasi-static during a short time slot (say, e.g., 1ms), we assume the channel remains constant within an OFDM symbol. In this letter, we explore Gaussian signaling for the JCASAR system, which achieves the capacity of point-to-point Gaussian channels \cite{b13}, with the achievable rate expressed as

\begin{equation}
C=B\sum_{i=0}^{N-1}{\log( 1+\frac{P_i\left| h_i \right|^2}{{\sigma _n}^2})}.
\end{equation}
\vspace{-3em}

\subsection{OFDM SAR Imaging Model}
Consider the JCASAR geometry shown in Fig. \ref{Fig1_geometry}. The UAV is moving parallelly to the y-axis and the instantaneous position of UAV can be expressed as $( 0,y_p\left( \eta \right) ,H_p)$, where $\eta$ denotes the slow time index in SAR imaging and $H_p$ is the altitude of the UAV. Given the bandwidth $B$, the range resolution of the system is $\rho _r=c/2B$. Specifically, we assume that there are $M$ range cells in the swath width. According to \cite{b7}, in order to achieve IRIC-free SAR imaging, the CP length should be at least $M-1$, resulting in $
T_{CP}=(M-1) /B$.

The received signal from the $m$-th range cell can be expressed as
\vspace{-1.9em}

\begin{equation}
\begin{aligned}
    u_m( t,\eta)& =g_m\varepsilon _a\left( \eta \right) \exp \{ -j4\pi f_c\frac{R_m\left( \eta \right)}{c} \}
\\             &\times \frac{1}{\sqrt{N}}\sum_{k=0}^{N-1}{S_k\exp \{ \frac{j2\pi k}{T}[ t-\frac{2R_m\left( \eta \right)}{c} ] \}}
\\             &+w\left( t,\eta \right) , t\in \big[ \frac{2R_m\left( \eta \right)}{c},\frac{2R_m\left( \eta \right)}{c}+T+T_{CP} \big],
\label{demodulated signal}
\end{aligned}
\end{equation}
where $R_m( \eta ) =\sqrt{\bar{R}_{m}^{2}+v_{p}^{2}\eta ^2}$, $
\bar{R}_m$ is the slant range when the UAV is closest to the center of the target, $g_m$ is the radar cross section (RCS) coefficient caused by the scatters in the $m$-th range cell, $\varepsilon _a( \eta )$ is the azimuth envelope, and $w$ denotes the noise. The complex envelope of the received signal from all the range cells can be written as
\begin{equation}
    y( t,\eta)=\sum_{m}u_m( t,\eta),
    \label{sum_signal}
\end{equation}
\vspace{-1em}

At the receiver, the baseband signal is sampled with sampling interval $
T_s=1/B$, in which case $t-{2R_m( \eta )}/{c}$ may be written as
\begin{equation}
t-\frac{2R_m\left( \eta \right)}{c}=t-\frac{2\left( R_n\left( \eta \right) +m\rho _r \right)}{c}
               =t-t_0\left( \eta \right) -mT_s
               \label{T after sampling},
\end{equation}
Let the sampling start at the moment when the received pulse begins, after a delay of $t_0\left( \eta \right)$ for the first arriving version of the transmitted pulse, combing \eqref{demodulated signal} - \eqref{T after sampling}, for a fixed slow time index, the received sequence may be recast as
\begin{equation}
y[i] =\sum_{m=0}^{M-1}{d_ms[i-m] +w[i]},\quad i=0, 1 ...N+2M-3 \label{{con:3}},
\end{equation}
where $d_m=g_m\varepsilon _a\left( \eta \right) \exp \{ -j4\pi f_c\frac{R_m\left( \eta \right)}{c} \} $ is the weighting RCS coefficient to be estimated, which contains the phase history within synthetic aperture time, $i$ denotes the fast time index, and $s[i-m]$ denotes the transmitted signal sampled at $\left(i-m\right)T_s$ in \eqref{transimtted signal} (i.e. $s[i-m]=s\left(\left(i-m\right)T_s\right)$). In this JCASAR system, we consider a special case called swath width matched pulse (SWMP) as elaborated in \cite{b7} (i.e. $M=N$), in which case we use $\boldsymbol{s}=[s_1, s_2, ... s_{N-1}, s_0, s_1, s_2, ... s_{N-1}]^T$ to denote the OFDM signal with CP in discrete time domain, and $[S_0, S_1, ...S_{N-1}]^T$ is FFT of the vector $[s_0, s_1, ...s_{N-1}]^T$.

Wiping off the first and last $M-1$ samples of the received sequence, the resulting sequence for range processing can be expressed as follows
\\
\begin{equation}
\boldsymbol{y}=d_0\left[ \begin{array}{c}
	s_0\\
	s_1\\
	\vdots\\
	s_{N-2}\\
	s_{N-1}\\
\end{array} \right] +d_1\left[ \begin{array}{c}
	s_{N-1}\\
	s_0\\
	\vdots\\
	s_{N-3}\\
	s_{N-2}\\
\end{array} \right] \cdots +d_{M-1}\left[ \begin{array}{c}
	s_1\\
	s_2\\
	\vdots\\
	s_{N-1}\\
	s_0\\
\end{array} \right] +\boldsymbol{w}.
\end{equation}
The matrix representation of (6) can be expressed as
\begin{align}
\boldsymbol{y}&=\left[ \begin{matrix}
	s_0&		s_{N-1}&		\cdots&		s_1\\
	s_1&		s_0&		\cdots&		s_2\\
	\vdots&		\vdots&		\ddots&		\vdots\\
	s_{N-1}&		s_{N-2}&		\cdots&		s_0\\
\end{matrix} \right] \,\,\left[ \begin{array}{c} 
	d_0\\
	d_1\\
	\vdots\\
	d_{N-1}\\
\end{array} \right] +\boldsymbol{w}
\\&=\boldsymbol{Sd}+\boldsymbol{w}.\nonumber
\end{align}

For SAR imaging application, the first step is to estimate the weighting RCS coefficient, followed by RCMC and azimuth processing to gegerate SAR image. Based on the linear signal model, the LS estimator for $\boldsymbol{d}$ reads \cite{b14}
\begin{equation}
\hat{\boldsymbol{d}}=(\boldsymbol{S}^H\boldsymbol{S})^{-1}\boldsymbol{S}^H\boldsymbol{y}.
\end{equation}
and the MSE of the estimator is $\text{tr}\big[\sigma ^2 (\boldsymbol{S}^H\boldsymbol{S} ) ^{-1} \big]$.

\section{Waveform Designs}
\subsection{Optimal Waveform Design for SAR Imaging}
The process of SAR imaging of the system can be summarized as range processing and azimuth processing. The weighting RCS coefficient is estimated through LS estimator from radar raw data, after RCMC, azimuth processing is executed on the weighting RCS by matched filtering. The matched filter depends on the velocity of the UAV, the waveform length, and the slant range, which can be expressed as
\begin{equation}
h_a=\exp \{ \frac{-j2\pi v_{p}^{2}}{\lambda R_c}t^2 \},
\end{equation}
which is independent of the transmitted signal. The optimization for SAR imaging application is to conceive a signal that is able to realize high-accuracy estimation of weighting RCS coefficients. Specifically, we use MSE to evaluate the accuracy of RCS estimation. As mentioned in section 2, the MSE of the proposed LS estimator is $\text{tr}\big[\sigma ^2 ( \boldsymbol{S}^H\boldsymbol{S} ) ^{-1} \big]$ . It should be noticed that, the matrix $ \boldsymbol{S}$ is a circulant matrix, which can be diagonalized as 
\begin{equation}
\boldsymbol{S}=\boldsymbol{F}^H\boldsymbol{\varLambda }_S\boldsymbol{F},
\end{equation}
where $\boldsymbol{F}$ is a $N\times N$ discrete Fourier transform (DFT) matrix and $\boldsymbol{\varLambda }_S$ represents diagonal matrix of symbols modulated on the OFDM subcarriers, in the form of $\boldsymbol{\varLambda }_S=\mathrm{diag}( \left[ S_0, S_1, \cdots S_{N-1} \right] ^T ) $. By doing so, the MSE of the LS estimator may be expressed in a compact form as
\begin{equation}
\begin{aligned}
&\text{tr}\big[ \sigma ^2( \boldsymbol{S}^H\boldsymbol{S} ) ^{-1} \big] 
\\
=&\sigma ^2\text{tr}\big[ \big( \boldsymbol{F}^H\boldsymbol{\varLambda }_{S}^{H}\boldsymbol{FF}^H\boldsymbol{\varLambda }_S\boldsymbol{F} \big) ^{-1} \big] 
\\
=&\sigma ^2\text{tr}( \boldsymbol{\varLambda }_{\left| S \right|^2}^{-1} ),
\end{aligned}
\end{equation}
where $\boldsymbol{\varLambda }_{\left| S \right|^2}$ is the diagonal matrix of the power on each subcarrier, which can be expressed as $
\boldsymbol{\varLambda }_{\left| S \right|^2}=\mathrm{diag}( \big[ \left| S_0 \right|^2, \left| S_1 \right|^2, \cdots \left| S_{N-1} \right|^2 \big] ^T )$. Accordingly, the optimal waveform design problem for SAR imaging may be formulated as
\begin{equation}
\begin{aligned}
&\min : \sigma ^2\text{tr}( \boldsymbol{\varLambda }_{\left| S \right|^2}^{-1} ) 
\\
&s.t.\sum_{i=0}^{N-1}{\left| S_i \right|^2}=P,\enspace \left| S_i \right|^2\geq 0.
\label{imaging only}
\end{aligned}
\end{equation}

Based on the analysis above, the optimal waveform for SAR imaging is the uniform power allocation while ensuring that the symbols modulated on each subcarrier maintain the same consistent modulus according to am-gm inequality.
\vspace{-0.5em}

\subsection{ISAC Waveform Design}
For JCASAR system, the signal should be random to convey communication information, so $S_k$ may not have a constant modulus, which jeopardizes the SAR imaging performance. To that end, we propose a JCASAR waveform design for random signals, while minimizing the MSE of LS estimator. In particular, we consider Gaussian signaling for our proposed JCASAR system, which is a typical assumption in the wireless communication literature, and is known to achive the capacity of Gaussian channels. We assume $S_k$ follows the complex Gaussian distribution, i.e., $S_k\sim \mathcal{C} \mathcal{N} \left( 0,P_k \right)$ where $P_k$ is the power allocated to the $k$-th subchannel. To account for the randomness of the signal, it is necessary to consider the expectation of the MSE (EMSE) of the LS estimator to evaluate the performance of SAR imaging. In order to achieve high accuracy SAR imaging while maintaining the performance of communication, the proposed ISAC waveform design minimizes the EMSE for the LS estimator under communication rate constraint, which is
\begin{equation}
\begin{aligned}
&\min :\mathbb{E} \big[ \sigma ^2\text{tr}( \boldsymbol{\varLambda }_{\left| S \right|^2}^{-1}) \big] 
\\
&s.t.   \sum_{i=0}^{N-1}{P_i}=P,\enspace    P_i\geq 0,
\\
& \quad \enspace \log\det ( \boldsymbol{I}+\boldsymbol{\varLambda }_P\boldsymbol{H} ) \geq R_0, \label{problem E}
\end{aligned}
\end{equation}
where $\boldsymbol{\varLambda }_P$ represents diagonal matrix of power allocated on the OFDM subcarriers, namely, $\boldsymbol{\varLambda }_P=\mathrm{diag}( \big[ P_0, P_1, \cdots P_{N-1} \big] ^T )$, and $\boldsymbol{H}$ represents diagonal matrix of the square of channel gain divided by the power of noise in a single OFDM symbol, given by $\boldsymbol{H}=\mathrm{diag}( \big[ \frac{\left| h_0 \right|^2}{\sigma ^2},\frac{\left| h_1 \right|^2}{\sigma ^2},\cdots ,\frac{\left| h_{N-1} \right|^2}{\sigma ^2} \big] ^T )$.

Due to the expectation operation in the objective function, the problem is difficult to solve directly. Recall that $S_k$ follows the complex Gaussian distribution, and each symbol on the subchannel is independent and identically distributed. The modulus of $S_k$ obeys the Rayleigh distribution, i.e., the probability density function of $
\left| S_k \right|$ can be expressed as $f_{\left| S_k \right|}( x ) =\frac{x}{P_k}\exp \{ -\frac{x^2}{2P_k} \}$. In that case, the EMSE of the LS estimator can be expressed as 
\begin{equation}
\begin{aligned}
&\mathbb{E} [ \sigma ^2\text{tr}( \boldsymbol{\varLambda }_{\left| S \right|^2}^{-1} ) ] \\ =&\sigma ^2\Bigg[ \mathbb{E} \{ \frac{1}{\left| S_0 \right|^2} \} +\mathbb{E} \{ \frac{1}{\left| S_1 \right|^2} \}+\cdots +\mathbb{E} \{ \frac{1}{\left| S_{N-1} \right|^2} \} \Bigg] 
\\
=&A\sigma ^2\text{tr}( \boldsymbol{\varLambda }_{P}^{-1}),
\end{aligned}
\end{equation}
where $A=
\int_{0^+}^{\infty}{\frac{1}{t}\exp \left\{ -t^2 \right\}}dt$.

The integral $\int_0^{\infty}{\frac{1}{t}\exp \left\{ -t^2 \right\}}dt$ is diverging because the integral function is singular at zero, which corresponds to $
\left| S_k \right|=0$ with probability density function $f_{\left| S_k \right|}=0$. Therefore we consider numerical integration within the integral interval that contains 99.9 $\%$ probability of $
\left| S_k \right|$. This will be used for the simulation in the next section. The power allocation problem in (\ref{problem E}) can be rewritten as a convex optimization problem

\begin{equation}
\begin{aligned}
&\min : A\sigma ^2\text{tr}( \boldsymbol{\varLambda }_{P}^{-1} ) 
\\
&s.t.  \sum_{i=0}^{N-1}{P_i}=P,\enspace P_i\geq 0,
\\
&\enspace \quad \log\det ( \boldsymbol{I}+\boldsymbol{\varLambda }_P\boldsymbol{H} ) \geq R_0.
\end{aligned}
\end{equation}

Compared with the MSE in problem (\ref{imaging only}), the EMSE is enlarged by a factor $A$, which is due to the randomness of the ISAC signal. Considering two special cases, one is that $R_0$ is small enough, or, equivalently, omitting the communication rate constraint, the solution of the problem is to allocate equivalent power among all subcarriers. We call this solution as imaging optimal power allocation. Another special case is the communication rate threshold $R_0$ is the biggest rate that the OFDM system can achieve, and the solution of this problem is known as water-filling solution: 
$P_k=( \frac{1}{\nu}-\frac{\sigma ^2}{\left| h_k \right|^2} ) ^+$. Any communication rate between the two extreme cases may be achieved by different power allocation strategies and the EMSE of estimator gradually increases as the communication rate $R_0$ increases, leading to a performance tradeoff between sensing and communications.
\section{Simulation Results}
In this section, we provide simulations and discussions for our proposed JCASAR system. The simulated SAR geometry is depicted in Fig. \ref{Fig1_geometry}. For computational efficiency, a fixed value positioned at the center of the range swath is established as the reference, which is a common practice in SAR imaging simulations. The azimuth processing procedure aligns with the Range-Doppler algorithm. The simulation experiments are performed with the following parameters: the height of UAV is \begin{math}
H_p=1 \text{km} \end{math}, the slant range swath center is \begin{math} R_c=\sqrt{2} \text{km} \end{math}, the velocity of UAV is \begin{math} v_p=40\text{m/s} \end{math} and the synthetic aperture time is \begin{math} T_a=1\text{s} \end{math}. The bandwidth of the ISAC signal is \begin{math} B=1.5\text{GHz} \end{math}, the number of subcarriers is \begin{math} N=64 \end{math}, the carrier frequency \begin{math} f_c=9\text{GHz} \end{math}, and the \begin{math} PFR=800\text{Hz} \end{math}. To evaluate the effect of the different signals on SAR imaging performance, we simulate both the SAR imaging result by ISAC signal (labeled as Gaussian) and OFDM with constant modulus $\left| S_k \right|$ (labeled as Constant modulus). 
\vspace{-1em}

\subsection{Point Target Examples}
\begin{figure}
\centering
\subfloat[]{\includegraphics[width=0.49\columnwidth]{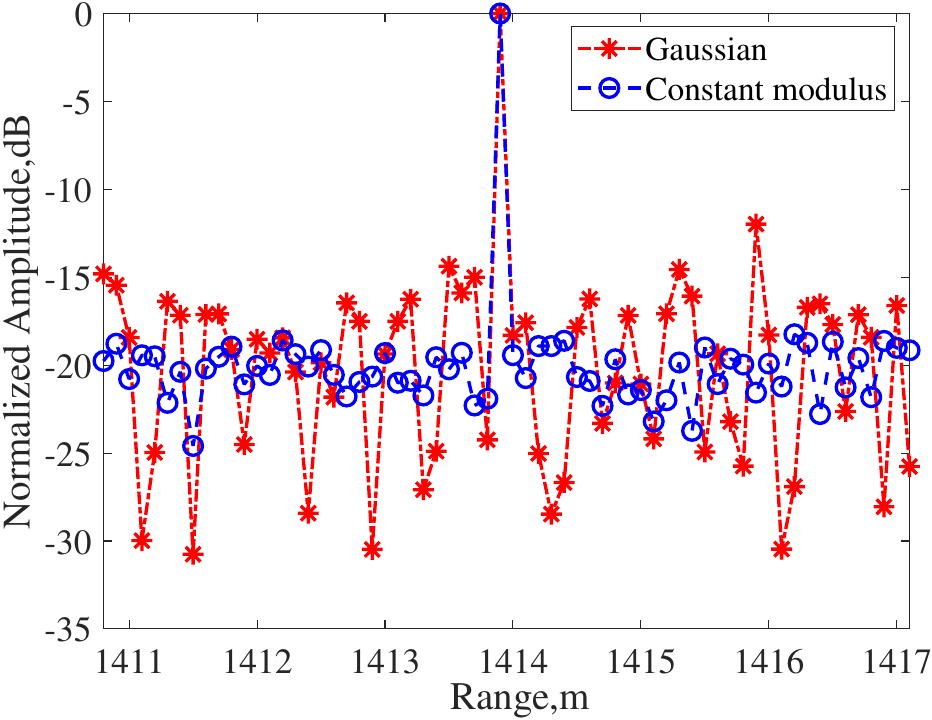}
	\label{range}}
\subfloat[]{\includegraphics[width=0.49\columnwidth]{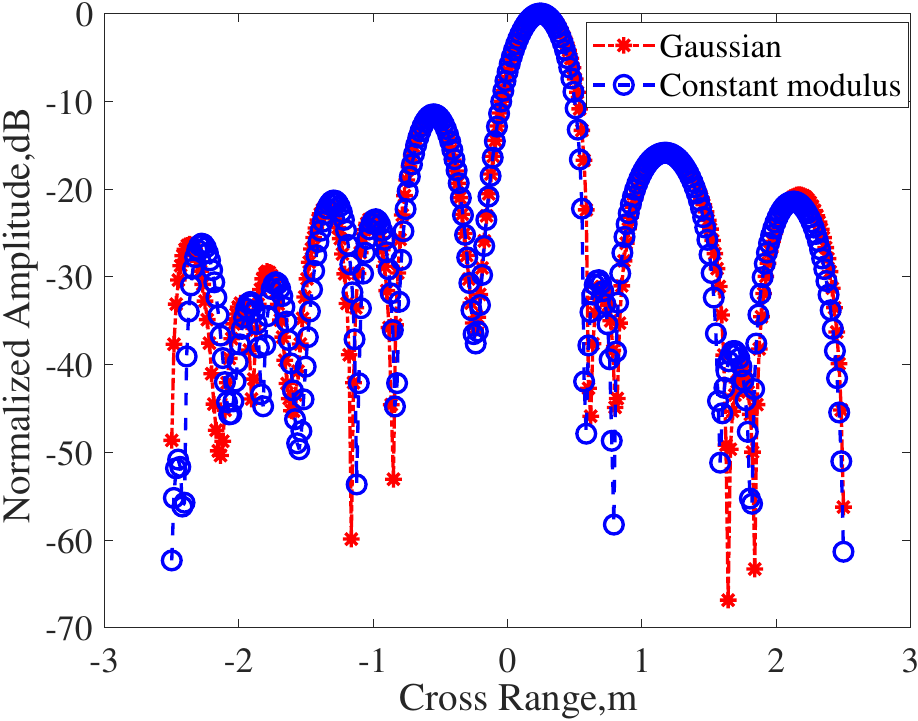}
	\label{azimuth}} \\
\subfloat[]{\includegraphics[width=0.49\columnwidth]{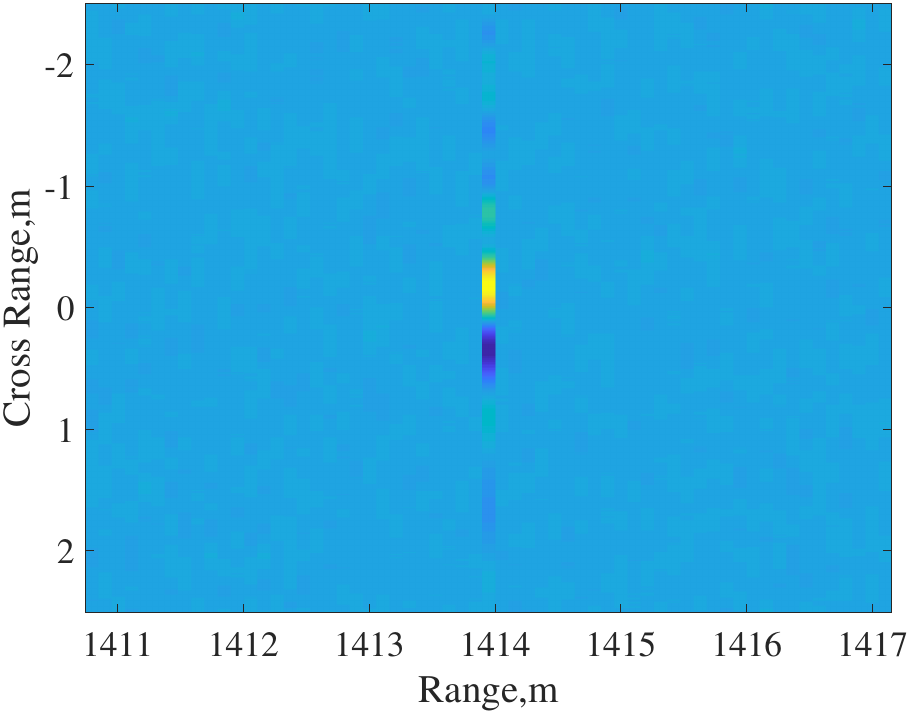}
	\label{point_constant}}
\subfloat[]{\includegraphics[width=0.49\columnwidth]{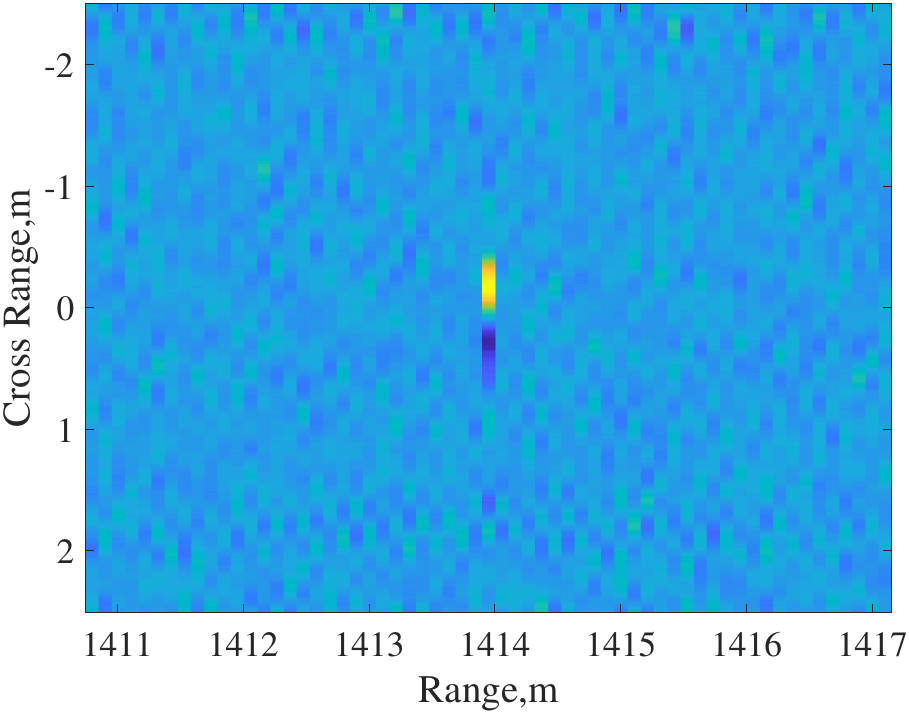}
	\label{point_Gaussian}}
\caption{SAR imaging of point target when SNR is 15dB. (a) Range profiles. (b) Azimuth profiles. (c) SAR image with constant modulus signal. (d) SAR image with Gaussian signal.}
\label{point_result}
\end{figure}

We commence by studying the case of a point target, positioned at the center of the range swath. The ISAC signal used in this simulation is the OFDM signal modulated by complex Gaussian distribution $S_k$ on each subcarrier, with uniform power allocation across subcarriers. The imaging results at an SNR of 15 dB are depicted in Fig. \ref{point_result}. Fig. \ref{range} and \ref{azimuth} illustrate the range and azimuth profiles of the SAR image with different signals. The range profile of the ISAC signal exhibits higher sidelobes compared to the constant modulus signal, resulting in a more blurred 2D SAR image as depicted in Fig. \ref{point_Gaussian}. However, it is noteworthy that the azimuth profiles with ISAC signal and constant modulus signal are nearly identical, as the azimuth processing is not directly related to the signal design.
\vspace{-1em}

\subsection{Extended Target Examples}
\begin{figure}
\centering
\subfloat[]{\includegraphics[width=0.48\columnwidth]{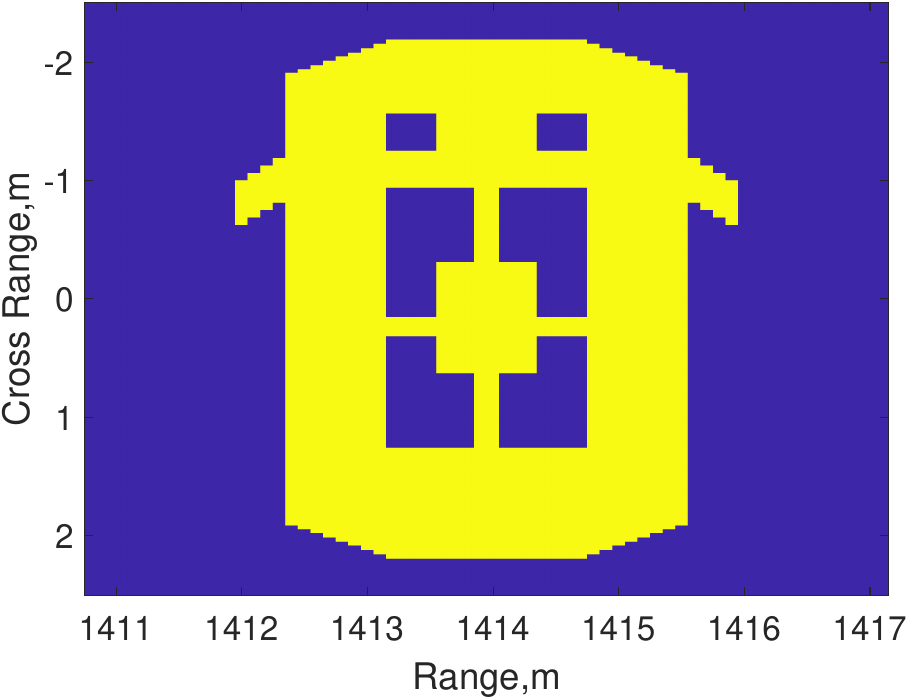}
	\label{origin}}
\subfloat[]{\includegraphics[width=0.48\columnwidth]{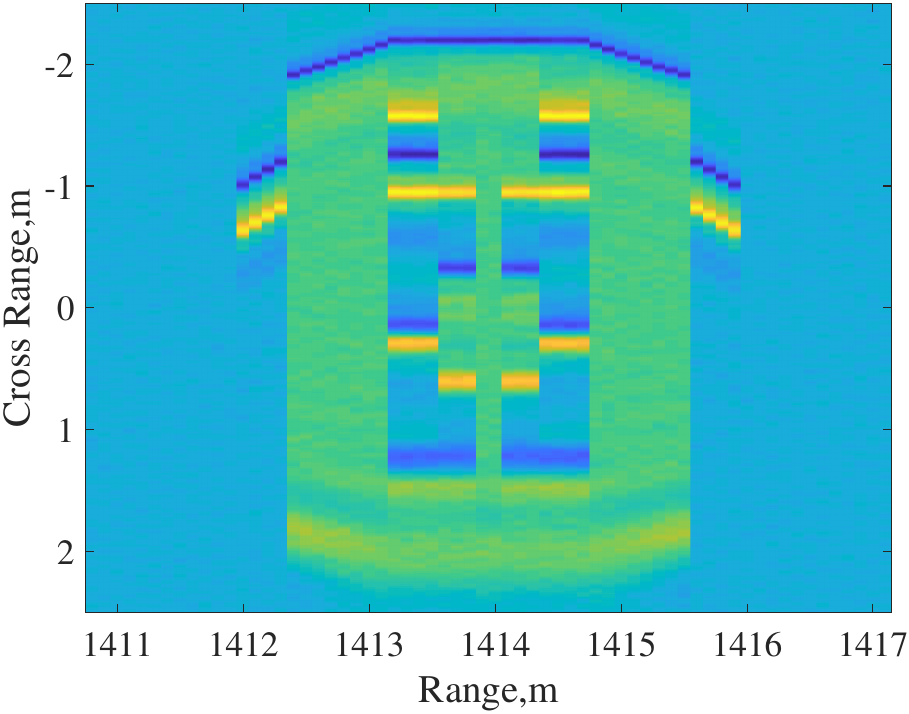}
	\label{extend_constant}} \\
\subfloat[]{\includegraphics[width=0.48\columnwidth]{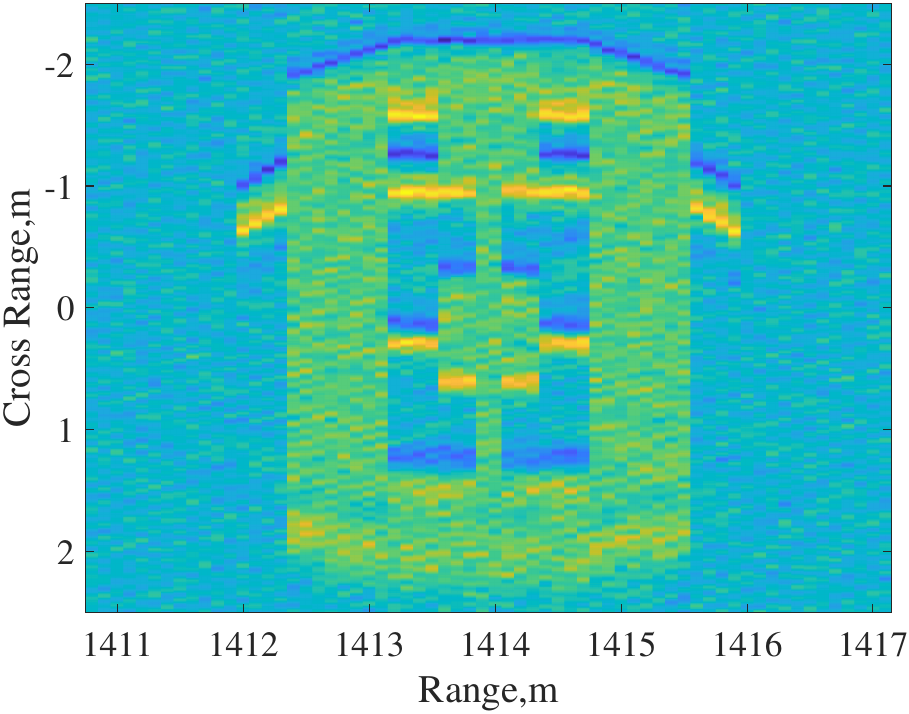}
	\label{extend_Gaussian}}
\subfloat[]{\includegraphics[width=0.48\columnwidth]{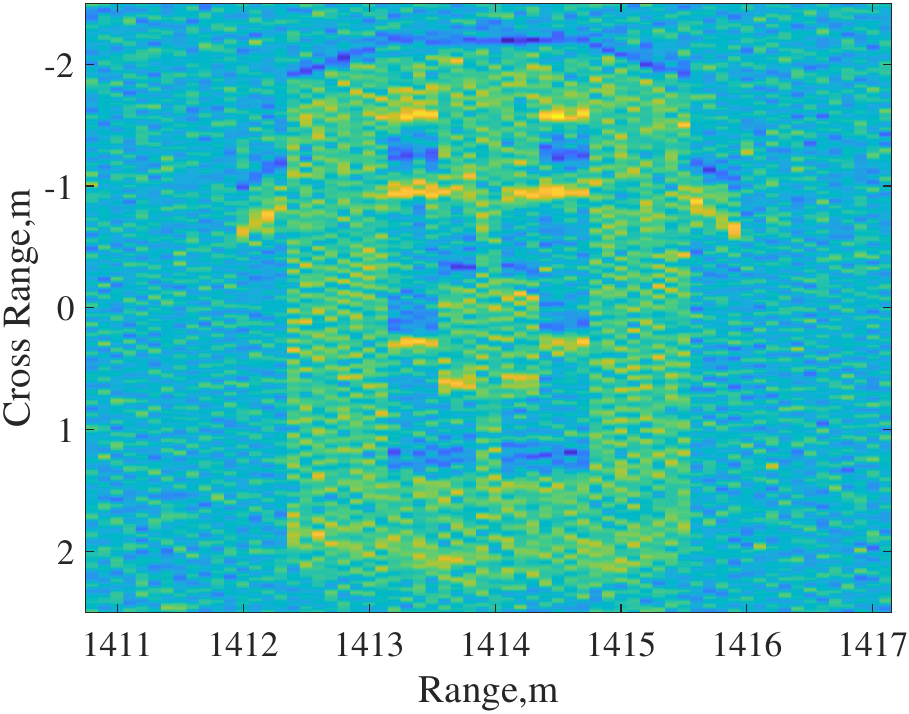}
	\label{extend_power_allocation}}
\caption{SAR imaging of extend target when SNR is 15dB. (a) Origin 2D image of car. (b) 2D SAR image with constant modulus signal. (c) 2D SAR image with Gaussian signal under uniform power allocation. (d) 2D SAR image with Gaussian signal under communication-optimal power allocation.}
\label{extend_result}
\end{figure}

For extended target, we examine the shape of a car modeled with several point scatters shown in Fig. \ref{origin}. Each line in the original image consists of 64 point scatter units, with the RCS set to one to represent the presence of the scatter and zero to indicate the absence of the scatter. The 2D SAR imaging result with the constant modulus signal at an SNR of 15 dB is shown in Fig. \ref{extend_constant}, which is clear enough to identify the shape of car. SAR imaging results with uniform power allocated Gaussian signal and communication optimal power allocated Gaussian signal are shown in Fig . \ref{extend_Gaussian} and Fig. \ref{extend_power_allocation}. Compared with imaging result with constant modulus signal, the images with Gaussian signal are more blurred, and the SAR image with communication optimal power allocated Gaussian signal is hardly to distinguish the shape of the car.

\vspace{-1em}

\subsection{Trade-off between SAR Imaging and Communication}
\begin{figure}[htbp]
\centering
\includegraphics[width=0.75\columnwidth]{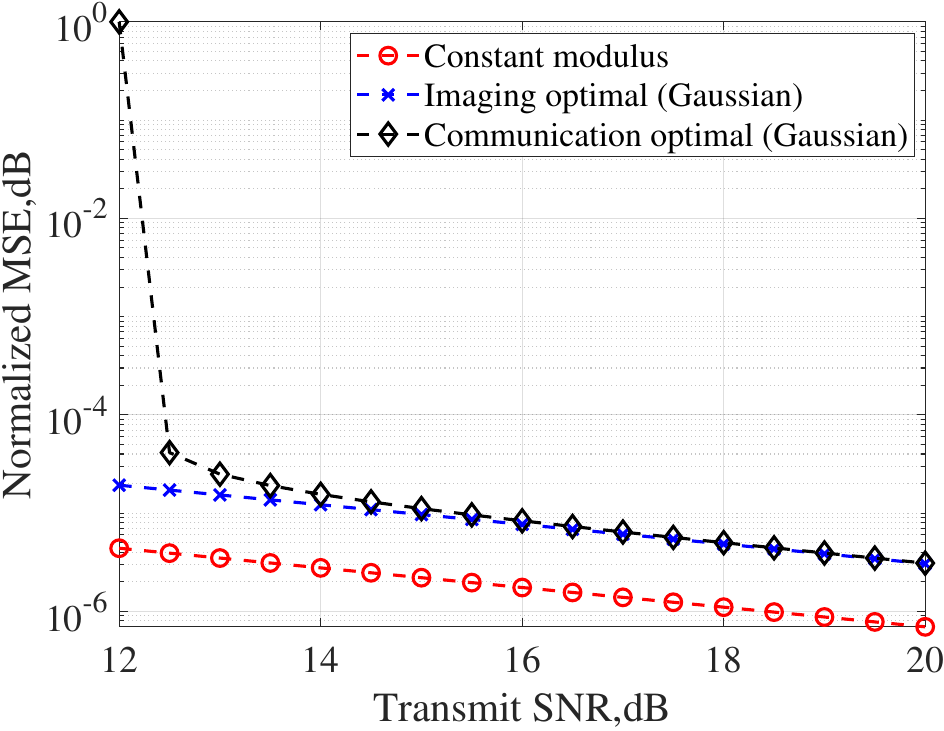}
\caption{Performance comparison between different signals.}
\label{MSE_VS_SNR}
\end{figure}

\begin{figure}[htbp]
\centering
\includegraphics[width=0.75\columnwidth]{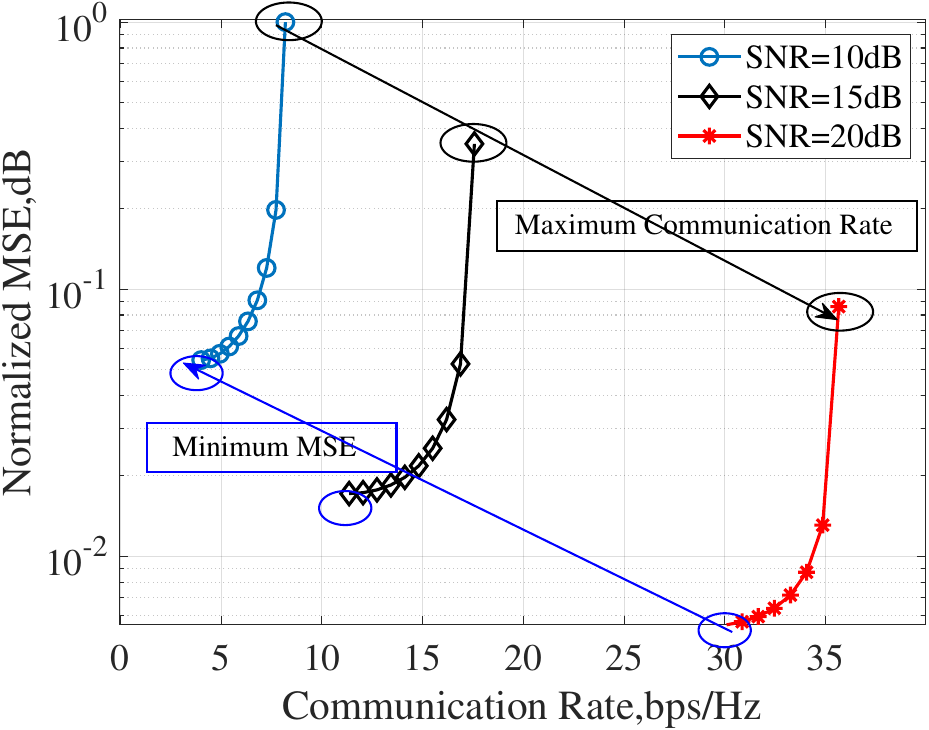}
\caption{Trade-off between SAR imaging and communication.}
\label{trade-off}
\end{figure}

Fig. \ref{MSE_VS_SNR} shows the normalized MSE of the LS estimator with different signal designs under different SNR. It is shown that, there is a gap in MSE with constant modulus signal and Gaussian signal as analyzed in the previous section. For different power allocation strategies, under low SNR, the MSE with communication optimal strategies is greater than their imaging-optimal counterparts, but the gap gradually dwindles when SNR is high. This phenomenon can be explained as that the water-filling solution tends to uniform distribution when the SNR is high.
Fig. \ref{trade-off} shows the MSE and communication rate of the ISAC signal under different SNR, which reveals the performance trade-off between SAR imaging and communication. As the figure illustrates, to achieve better performance of SAR imaging, communication rate has to be sacrificed. More importantly, a scalable tradeoff between imaging optimal and communication optimal performance can be achieved through different power allocation strategies for ISAC signal designs.

\vspace{-1em}
\section{Conclusion}
This letter presents a joint communication and SAR imaging system, which is capable of reconstructing the target profile while serving a communication user. Based on the MSE of the LS estimator, we propose the optimal waveform design for SAR imaging only application and the power allocation strategies for the JCASAR task. Performance of SAR imaging with different signal designs is validated through numerical simulation and finally we reveal the performance trade-off between SAR imaging and communication. This approach offers a novel perspective on utilizing random signals for both target imaging and communication purposes.
\vspace{-1em}

\bibliographystyle{IEEEtran}   
\bibliography{ref}          
\end{document}